\documentclass[11pt]{article}
\usepackage{blois,epsfig}

\newcommand{\href}[2]{ \, #2}

\begin{document}

\title{Average transverse momenta in hyperon production at p-p collider experiments}

\author{Olga Piskounova}

\address{P.N.Lebedev Physics Institute of Russian Academy of Science, Leninski prosp. 53, 119991 Moscow, Russia}

\maketitle\abstracts{The previously publicated analysis of transverse momentum spectra of $\Lambda^0$ hyperons from
LHC experiments (ALICE, ATLAS, CMS)in the comparison with earlier experiments was reconsidered with correct spectra from STAR collaboration. The LHC data at $\sqrt{s}$ = 0.9 and 7 TeV and the data of proton-proton experiments of lower energies were fitted with the universal formula that includes the energy dependent slope as the main parameter. The dependence of average transverse momenta  on $\sqrt{s}$ has been obtained with the help of this formula. The asymptotics of the energy dependence of $<p_t>$  shows the behavior $ ~ s^{0.05}$, that was not expected in early description of hadron transverse momentum in the framework of Quark-Gluon String Model. The previous important conclusion about spectra of cosmic rays was not changed: the long debated "knee" in the cosmic proton spectra at $E_p= (2,5 - 4)*10^{15}$ eV in laboratory system can not be considered any more as the result of dramatic changes in the dynamics of baryon hadroproduction. The reason of the steadily growing of $<p_t>$ seems situated outside the predictive power of QGSM. Nevertheless the average transverse momentum can reach a constant value with higher energies that has been predicted in our model long ago.}

\section{Introduction}

As you will see from the reference list to this article, my study is based on the papers of general authors of the Quark-Gluon String Model: Prof.A.B.Kaidalov and Prof.K.A.Ter-Martirosyan \cite{classicpapers}.
Our collaboration began in early 1980th, while it was necessary to predict proton and pion spectra in the whole
kinematical regions (including the arrea of beam particle fragmentation) that was caused by the problems in cosmic ray physics. Practically, only quark-gluon strings could help to account correctly the energy conservation between different regions of hadron spectra. The entire corpus of the data on hadron spectra that have been measured at collider experiments was described those times as well as the cross sections for the heavy quark production was predicted. In 90th, when it became clear that the ratios of spectra can help to reveal the quark structure of interacting particles, we have studied all possible effects of leading/nonleading asymmetries, which are caused by
different quark contents of interacting hadrons. Only one subject of QGSM application has left non-upgraded with new data, it is the description of transverse momentum spectra of produced particles. So, now I continue this activity on the new level of collider investigations.     

The transverse momentum spectra of $\Lambda^0$ hyperons that were measured in many hadron collider experiments: ISR \cite{isr}, STAR \cite{star}, UA1 \cite{ua1}, UA5 \cite{ua5}, CDF \cite{cdf} and experiments at LHC \cite{alice,atlas,cms}. The given comparison shows the different forms of distributions that are changing from lower energy to higher. As it was established in the previous publication \cite{antiproton}, there is specific dependence of spectra on the type of colliding hadrons. Due to this difference, we have to confine ourself to the consideration only of the proton-proton collision results : ISR, STAR, ALICE, ATLAS and CMS in the wide range of $\sqrt{s}$: beginning from 53 GeV to LHC energy 7 TeV. 

This range of energy has motivated me to investigate the behaviors of baryon spectra, because the proton spectrum in cosmic rays (CR)  shows the "knee" at the very energy between Tevatron and LHC experiments \cite{knee}, see Fig.1. The bent slope of proton spectrum may be of the astrophysical origin otherwise it means a substantial change in dynamics of particle production. We have to study the behaviors of baryon production at the collider experiments from one energy to another in order to conclude about whether or not the dramatic changes have place in baryon production processes at the "knee" energy and a little above.
 
\begin{figure}[thb]
\begin{center}
\epsfig{figure=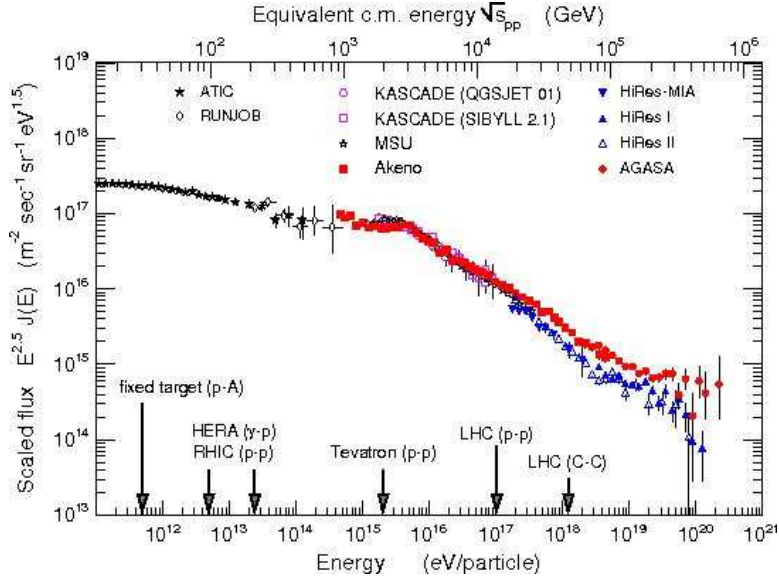,height=3.0in}
\caption{The cosmic proton spectrum at the energies between Tevatron and LHC.}
\label{fig:knee}
\end{center}
\end{figure}

The average transverse momenta of hyperons can show the specifical points in the characteristics of baryon production especially in the considered range of energies. As it was learned from the previous QGSM studies, the typical average transverse momentum is almost constant or it can slightly grow with energy due to growing contributions of multipomeron exchanges that can give larger fluctuations in transverse momenta.

\section{Baryon transverse momentum distributions at LHC and QGSM description.}

The recent data on $\Lambda^0$ hyperon distributions are obtained in the following LHC
groups: ALICE \cite{alice} at 900 GeV, ATLAS \cite{atlas} at 900 GeV and 7 TeV and CMS \cite{cms} at 900 GeV and 7 TeV. The lower energy experiments, which we can compare with the results of LHC groups, are ISR \cite{isr}($\sqrt{s}= 53 GeV$) and STAR \cite{star}($\sqrt{s}= 200 GeV$).

If we fit the data with a simple exponential function $e^{-B*p_t}$, both experiments, ATLAS and CMS, have presented the spectra at 7 TeV with the similar slopes, see Fig.2. 

\begin{figure}[thb]
\begin{center}
\epsfig{figure=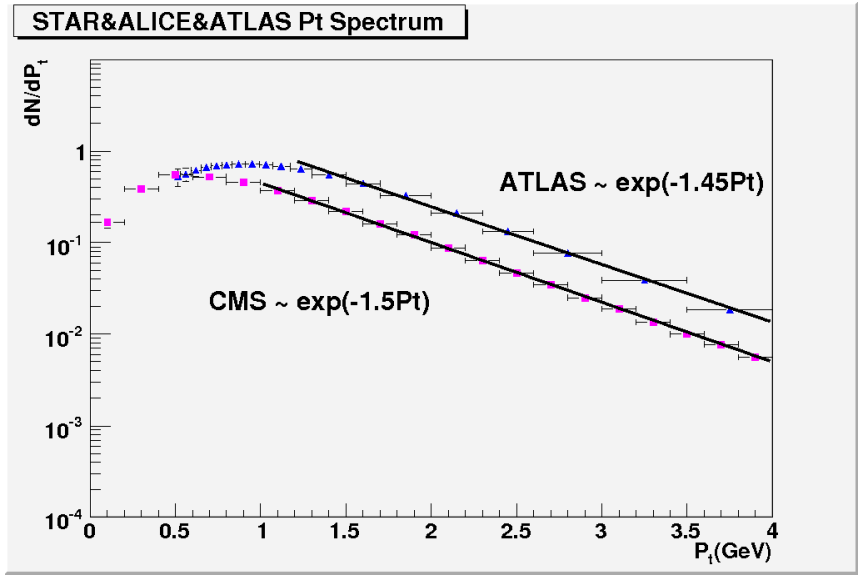,height=3.0in}
\caption{The comparison of ATLAS and CMS data at $\sqrt{s}=7TeV$.}
\label{fig:atlas}
\end{center}
\end{figure}

The different form of the distributions at low $p_t$'s may be caused, on my mind, by some specifics in the efficiency of the detecting procedure. As we see here, ATLAS has systematically lower efficiency for $\Lambda^0$  at $p_t< 1$ GeV in comparison to the results of other LHC groups.

It was studied in our early QGSM paper \cite{veselov} that the transverse momentum spectra of hadrons after proton-proton collisions can be perfectly described with a bit more complicate dependence:

\begin{equation}
\frac{dN^{H}}{dp_{t}} \propto  p_{t}*E \frac{d^{3}\sigma^H}{dx_F d^{2}p_{t}} \propto p_{t}*e^{-B_0*(m_t-m_0)},
\end{equation}

where $m_0$ is the mass of produced hadron, $m_t$ = $\sqrt{p_t^2+m_0^2}$ and $B_0$ used to bring the dependence on $x_F$, but in central region of rapidity this slope is taken constant. The slopes of the spectra of many types of hadrons ($\pi$, K, p and $\Lambda^0$) were estimated for the data of early proton-proton collision experiments of the energies that were available those times. The slopes of baryon spectra was approximately $B_0$ = 6,0 which has been expected to be stable at all energies.
Now we have to conclude that $B_0$ depends strongly on the energy of proton interaction. More, as it is seen from the spectra at LHC and RHIC, the value of $m_0$ is not equal nor to proton nor to hyperon masses.

\begin{figure}[thb]
\begin{center}
\epsfig{figure=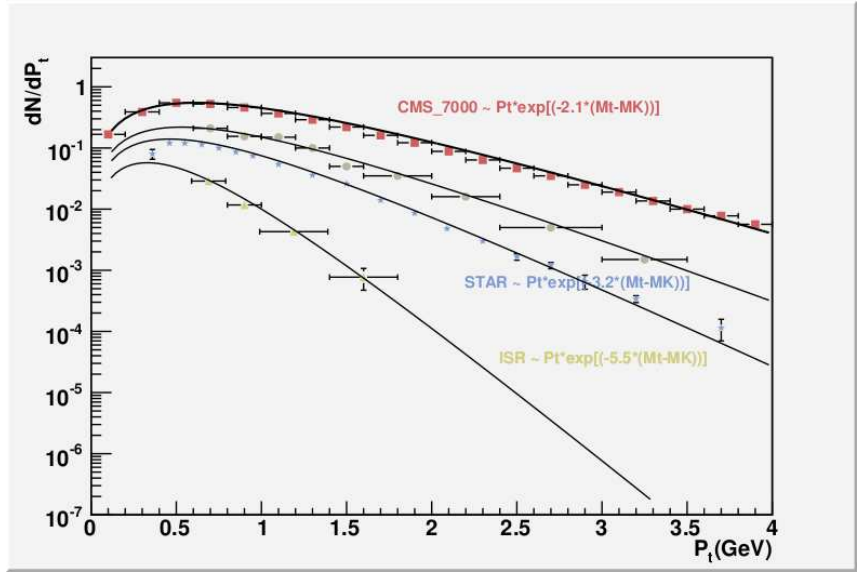,height=3.0in}
\caption{The description of STAR, ALICE and CMS data with the QGSM function.}
\label{fig:STALCMS}
\end{center}
\end{figure}

The better description of hyperon spectra, which is shown in Fig.3, can be achieved  with $m_0$ = 0,5 GeV that is kaon mass. This fact can be explained by the minimal quark-antiquark chain at hyperon production, which is consisted of $K+\Lambda^0$ (only in the case of proton-proton collision) and the minimal transverse momentum of $\Lambda$, such a way, should be of order of kaon mass. In suggested fit, the values of slope parameter have to be equal for all LHC experiments of the same energy. It means that as soon as we estimate the slopes for various collider energies we are getting a chance to calculate (using the formula above) the energy dependence of average transverse momentum for the wide range of proton-proton experiments.

\section{Average transverse momenta at $\sqrt{s}$ = 53, 200 GeV, 900 GeV and 7 TeV}

The resulting dependence of average $p_t$ on the energy of interaction is shown in Fig.4.
After QGSM study of transwerse momentum spectra of hadroproduction \cite{veselov} that was carried out almost three decades ago, it seems reasonable to expect no energy dependence of the average $p_t$'s. Nevertheless $<p_t>$ is growing, but the stable slight growing of the average $p_t$ at the energies of LHC may indicate the constant average
$p_t$, which can be reached at futher measurements with higher energies. In this sence the old QGSM predictions
will be considered beeing true. 

\begin{figure}[thb]
\begin{center}
\epsfig{figure=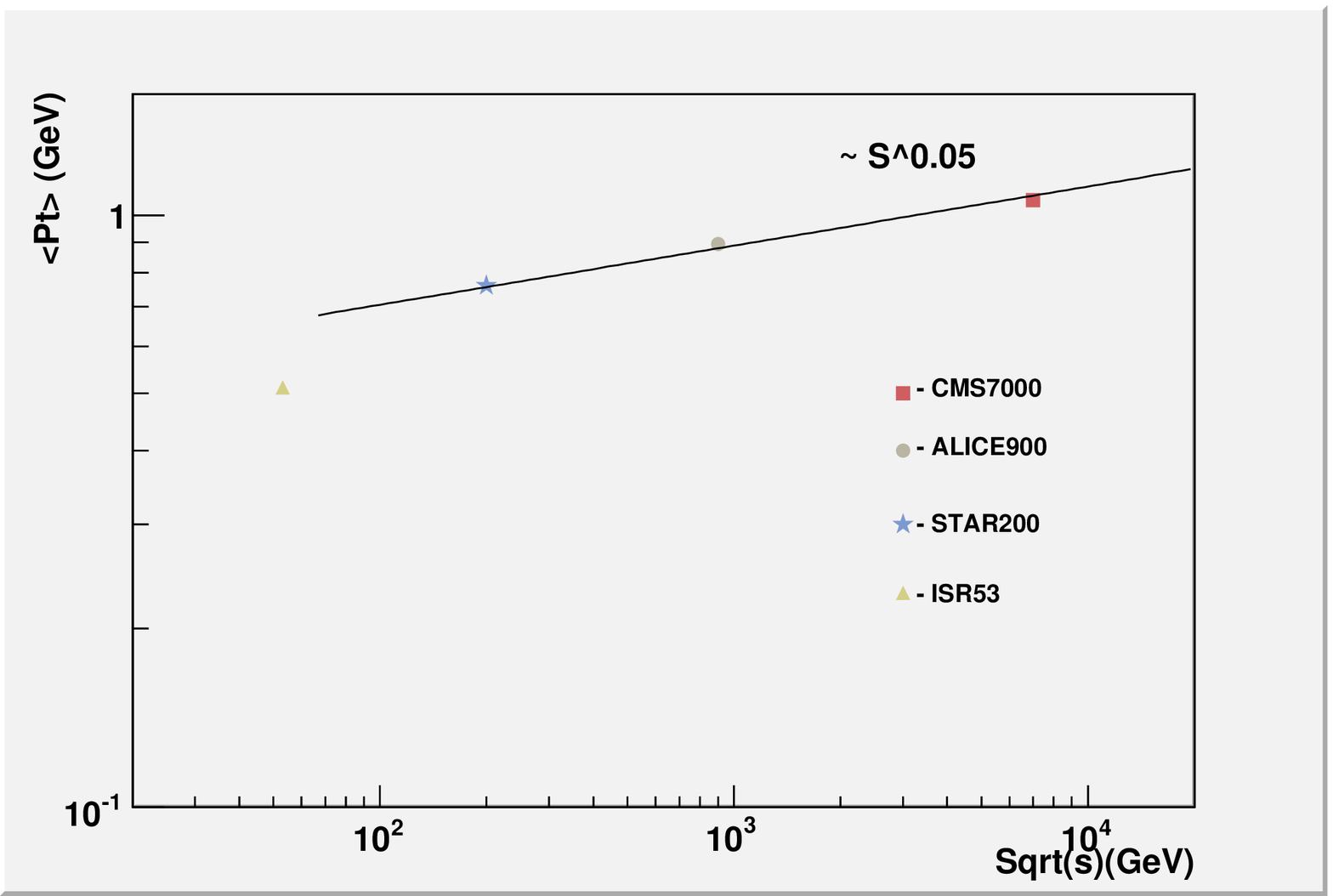,height=3.0in}
\caption{The average transverse momentum of hyperons vs. the energy of colliding protons.}
\label{fig:ptavrg}
\end{center}
\end{figure}

We could only suggest that the fast growing of $<p_t>$ before STAR energy is due to valuable impact of the baryon production from fragmentation region of proton. For higher energy, more than $\sqrt{s}$ = 200 GeV, the leading production of $\Lambda^0$ doesn't already play any important role.

\section{Conclusions}

The review of results on transverse momentum distributions of hyperons that are produced in proton-proton collisions of up-to-date energies reveales a significant change in the slopes of spectra in the region of $p_t$ = 0,3 - 4,0 GeV/c. The spectra of baryons are becoming harder and harder with the energy growth from ISR ($\sqrt{s}$=53 GeV) to RHIC ($\sqrt{s}$=200 GeV) and LHC (0,9 and 7 TeV). The detailed analysis of hyperon spectra, which is analogous to our early studies in the frameworks of Quark-Gluon String Model, demonstrates the change of slopes from $B_0$ = 5,5 (ISR) to $B_0$ = 2,1 (LHC at 7 TeV). 

As the result, the average $<p_t>$ value is growing up to approximately $\sqrt{s}$= 200 GeV and then it goes with the asymptotics ~ $s^{0.05}$. This behavior can not be considered as important change in hadroproduction processes. This statement is very important for cosmic ray physics, where the "knee" (the change in the slope) at $E_{lab}$ $\approx$ 3* $10^{15}$ eV in cosmic proton spectra might have an origin in hadronic interactions. As we have discussed above, nothing drastic happens with baryon spectra up to $\sqrt{s}$ = 7 TeV that corresponds to $E_{lab}= 2,5*10^{16}$ eV. It means that the "knee" can be caused  only by an astrophysical reason. On the other hand, the "knee" may indicate, as an example, the maximal energy of protons that are beeing produced in other Galaxies. But the idea of production of very high energy protons in space assumes a futher detailed investigations of the production dynamics of quark systems in the framework of our model. It also seems very intriguing to observe whether or not the average transverse momentum of baryons becomes a constant at futher energies of LHC. 

Some brief analysis of growing-with-energy antiparticle-to-particle ratious of elementary particles that are measured in cosmic experiments was done in the QGSM technics \cite{aptopratio}. The growing ratios may be explained by the leading behavior of hadron production spectra that is inavitable result of "positiveness" of our Universe. The antimeson-to-meson ratios in cosmic ray physics as well as antibaryon/baryon production asymmetries, which are already measured in LHC experiments, are intended to be discussed in the upcoming publication.
 
\section{Acknowledgments} 

I have to express my gratitude to the noncommercial organizations, Russian Foundation of Basic Research (grant №  13-02-06091) and Dmitry Zimin's "Dynasty" foundation, whoes financial supports made possible the First Kaidalov's Phenomenology Workshop. I hope as well that Kaidalov's contributions into the phenomenology of QCD physics will give more and more results in the explanation of the contemporary experimental data.

\end{document}